# Evaluating the Evolution of Critical Thinking, Creativity, Communication and Collaboration in Higher Education Courses

Margarida Romero[1,2]
[1]Université Côte d'Azur, Smart EdTech, Nice, France.
[2]IIIA, CSIC, Barcelona, Spain.

## Abstract

The development of Creativity, Communication, Critical Thinking, and Collaboration (the 4Cs) is a central objective of contemporary competency-based education. However, empirical evidence on how these competencies evolve across learning modules and instructional phases remains limited. This study evaluates the evolution of the 4Cs from pre-pilot to pilot implementation phases across three educational contexts, using the project's 4Cs theoretical framework as an analytical lens. The analysis of three pilot cases (IASIS, EASD, and UPATRAS) compares the 4Cs scores to identify patterns of growth, stagnation, or decline over time. Results indicate that communication and critical thinking showed the most consistent and substantial improvements, particularly in pilots with lower pre-pilot baselines, suggesting that structured pilot interventions effectively support cognitive and expressive competencies. In contrast, creativity exhibited context-dependent outcomes, while collaboration emerged as the most fragile competency, often stagnating or declining during scale-up. Interpreted through the theoretical framework, these findings suggest that competency evolution is strongly shaped by instructional design, assessment alignment, and learning activity structures rather than learner ability alone. The study contributes empirical validation to the 4Cs framework and highlights the need for differentiated, competency-sensitive design and evaluation strategies when scaling educational modules.

## Keywords

# 1. Introduction

The prominence of transversal competencies has grown in educational research, policy, and practice. Frameworks developed by international organisations consistently emphasise the importance of competencies that extend beyond disciplinary knowledge and enable adaptation, innovation, and effective participation in complex environments (Cosgrove & Cachia, 2025; OECD, 2019;). Among these, Creativity, Communication, Critical Thinking, and Collaboration, commonly referred to as the 4Cs, are recognised as core competencies underpinning learning, employability, and lifelong development (Dumas & Kaufman, 2024; Romero, 2015). Within competency-based education, an important distinction is made between competence as observable performance and competency as the level of proficiency with which that performance is enacted (Argyris & Schön, 1974; Voorhees, 2001). Skills are thus considered components of broader competencies, which are context-sensitive and evolve through practice rather than remaining static attributes (Westera, 2001). Nevertheless, the assessment of transversal competencies remains methodologically challenging (Ananiadou & Claro, 2009). These competencies are multidimensional, domain-sensitive, and often manifested through complex learning activities rather than discrete outcomes. Traditional assessment approaches may therefore struggle to capture their full scope, particularly when competencies develop gradually and unevenly across instructional phases and require consideration of contextual factors (Brianza et al., 2024; Duret & Romero, 2025). Rubric-based assessment addresses these challenges and allows competencies to be operationalised through clearly defined criteria and levels of achievement, supporting alignment between learning objectives, instructional design, and assessment practices. Beyond their evaluative function, rubrics promote transparency, learner self-regulation, and shared understanding of quality between educators and learners (Panadero & Romero, 2014; Panadero et al., 2023). They are particularly well suited to competency-oriented learning environments, where assessment must accommodate both formative and summative purposes, while supporting longitudinal analysis of competency development.

## 2. 4Cs Dynamics

Each of the 4Cs is driven by distinct pedagogical mechanisms and cognitive processes. Critical thinking is commonly associated with problem identification, interpretation, exploration, and analysis, drawing on epistemic judgement and reflective reasoning (Facione, 1989, 2011; Garrison et al., 2001; Halpern, 2006). Creativity is typically defined through the interplay of novelty and value, encompassing divergent and convergent thinking processes that lead to solutions that are original, effective, and implementable within a given context (Amabile et al., 2018; Guilford, 1967) Communication develops through opportunities for expression, interaction, and feedback, while collaboration is particularly sensitive to task design, reciprocity, and the quality of social interaction, often proving more fragile under conditions of scale or reduced synchrony (Jeong & Hmelo-Silver, 2016). Transitions between instructional phases may therefore differentially support or constrain specific competencies. Increased standardisation and assessment alignment may favour cognitive and expressive competencies, such as critical thinking and communication, while limiting opportunities for creative exploration or sustained collaborative engagement.

The present study investigates the evolution of the 4Cs across learning modules over time. By analysing longitudinal assessment data collected across distinct implementation phases, the study aims to identify patterns of growth, stability, or decline in each competency and to interpret these patterns in relation to established theoretical perspectives on competency development and assessment. In doing so, the study contributes evidence-based insights into how transversal competencies evolve in modular learning contexts and informs the design of learning environments that seek to foster balanced and sustainable development of the 4Cs.

## 3. Methodology

This study adopted a longitudinal, quasi-experimental design to examine the development of the 4Cs (Creativity, Communication, Critical Thinking, and Collaboration) across the three pilot course implementations. Competency development was analysed by comparing learner outcomes between a pre-pilot phase and a subsequent pilot phase within three institutional contexts: IASIS, Environmental Ambassadors for Sustainable Development (EASD), and the University of Patras (UPATRAS). To support explanatory interpretation of observed changes, competency data were analysed in relation to both academic performance indicators and learner activity logs, allowing the study to address not only whether competencies developed, but also why such changes may have occurred.

### 3.1. Participants

Participants were learners enrolled in courses delivered within the three pilot institutions. Although all pilots were guided by a shared pedagogical framework, local adaptations were made in instructional design and delivery to reflect institutional contexts. For the IASIS pilot, the dataset includes 255 participants, while the IASIS pre-pilot phase involved 67 participants. In the EASD pilot, a total of 130 participants were recorded, compared to 118 participants in the pre-pilot phase. For UPATRAS, competency assessment data were available for both pre-pilot and pilot phases, while grade data were available only for the pilot phase, which involved 251 participants. All datasets were anonymised prior to analysis and collected in accordance with institutional ethical standards (**Table 1**).

**Table 1.** Participants in the pilot and pre-pilot.

| Pilot | Pre-pilot (n) | Pilot (n) |
|---|---|---|
| IASIS | 67 | 255 |
| EASD | 118 | 130 |
| UPATRAS |  | 251 |

### 3.2. Procedure

Learner competencies were measured using structured rubric-based assessment instruments aligned with the Horizon augMENTOR project's 4Cs theoretical framework

([Septiani et al., 2023](#)). These instruments operationalised Creativity, Communication, Critical Thinking, and Collaboration through predefined criteria and performance levels, producing numerical scores on a standardised scale for each competency. The rubrics were integrated into the assessment practices of each pilot context and applied consistently across pre-pilot and pilot phases. The instruments were theoretically grounded in established definitions of the 4Cs and had been previously applied. Internal validation procedures were conducted, including data cleaning, consistency checks, and score standardisation across pilots to ensure comparability of measurements. Each competency was represented by a numerical score on a standardised scale and aggregated at the learner level. Prior to analysis, all datasets underwent cleaning and validation procedures to remove incomplete records and ensure consistency across pilots. Variable names and scoring formats were standardised to enable cross-pilot comparison, and competency scores were aggregated to compute mean values per learner and per phase. Competency, grade, and activity log datasets were then integrated using anonymised learner identifiers. Only learners with valid and matching data across the relevant datasets were included in each analytical step. Longitudinal analysis was conducted to examine changes in the 4Cs between the pre-pilot and pilot phases, except for University of Patras. Mean competency scores were calculated for each phase and compared within each pilot to identify absolute changes over time. This analysis enabled the identification of competencies that improved, remained stable, or declined during pilot implementation, as well as differences in developmental patterns across institutional contexts.

Learner activity logs were analysed using engagement metrics including activity completion frequency, number of interactions within learning activities (e.g., submissions or participatory actions), and time-on-task. These indicators were examined in relation to changes in 4C scores at the pilot level to identify behavioural patterns associated with observed competency gains or declines.

## 4. Results

This section presents the results of the longitudinal analysis of the 4Cs across pre-pilot and pilot implementation phases. Competency scores are reported as mean values aggregated at module level for each pilot context. **Table 2** presents the aggregated scores for all the pilots, considering the mean as well as the standard deviation.

In the following sections, changes over time are expressed as the difference between pilot and pre-pilot mean scores (Δ).

### 4.1. Evolution of the 4Cs by Pilot Context

In the IASIS pilot, creativity increased from the pre-pilot phase (M = 70.27, SD =

**Table 2.** Participants in the pilot and pre-pilot.

| Pilot | Phase | Creativity | Communication | Critical thinking | Collaboration |
|---|---|---|---|---|---|
| | | M (SD) | M (SD) | M (SD) | M (SD) |
| IASIS | Pre-pilot | 70.27 (25.88) | 64.23 (22.95) | 69.66 (21.92) | 78.85 (18.34) |
| | Pilot | 72.80 (19.57) | 75.77 (16.67) | 75.38 (17.09) | 76.87 (16.66) |
| EASD | Pre-pilot | 80.42 (18.41) | 75.01 (23.57) | 81.05 (18.02) | 82.60 (16.26) |
| | Pilot | 76.05 (20.78) | 80.18 (21.07) | 77.35 (21.51) | 78.45 (20.93) |
| UPATRAS | Pre-pilot | 74.44 (35.58) | — | 66.10 (24.65) | 79.46 (12.84) |
| | Pilot | 81.49 (31.37) | — | 95.09 (10.18) | 79.96 (33.92) |

25.88) to the pilot phase (M = 72.80, SD = 19.57; Δ = +2.53). Communication showed the largest improvement, increasing from M = 64.23 (SD = 22.95) to M = 75.77 (SD = 16.67; Δ = +11.54). Critical thinking also increased from M = 69.66 (SD = 21.92) to M = 75.38 (SD = 17.09; Δ = +5.72), whereas collaboration showed a small decline from M = 78.85 (SD = 18.34) to M = 76.87 (SD = 16.66; Δ = −1.98).

In the EASD pilot, communication increased from M = 75.01 (SD = 23.57) in the pre-pilot phase to M = 80.18 (SD = 21.07; Δ = +5.17). In contrast, decreases were observed for creativity (Δ = −4.37), critical thinking (Δ = −3.70), and collaboration (Δ = −4.15).

In the UPATRAS pilot, substantial gains were observed in critical thinking, which increased from M = 66.10 (SD = 24.65) in the pre-pilot phase to M = 95.09 (SD = 10.18; Δ = +28.99). Creativity also increased (Δ = +7.05), while collaboration remained largely stable (Δ = +0.50).

## 4.2. Cross-Pilot Comparison of Competency Evolution

Across pilots, communication showed consistent improvement in all contexts where data were available (IASIS and EASD). Critical thinking exhibited the greatest variability, ranging from moderate gains (IASIS) to substantial gains (UPATRAS) and moderate declines (EASD). Creativity outcomes varied by context, with increases observed in IASIS and UPATRAS and a decline in EASD. Collaboration showed the least positive development overall, with decreases observed in IASIS and EASD and only a negligible increase in UPATRAS.

# 5. Discussion

This study examined the evolution of the 4Cs competencies (Creativity, Communication, Critical Thinking, and Collaboration) across learning modules from pre-pilot to pilot implementation phases. The results reveal that the four competencies did not evolve uniformly, either within or across pilot contexts. Instead, distinct patterns emerged that reflect the differential sensitivity of each competency to instructional design, assessment alignment, and conditions of scale.

## 5.1. Differential Evolution of the 4Cs

One of the most salient findings is the consistent improvement of communication in all pilot contexts where data were available. In both IASIS and EASD, communication showed clear gains despite divergent trends in the other competencies. This pattern aligns with theoretical perspectives that position communication as a competency that benefits strongly from increased structure, explicit criteria, and frequent opportunities for practice and feedback. As communication skills are often directly embedded in assessed tasks, such as written assignments, presentations, or structured interactions, they are likely to be reinforced during pilot phases characterised by clearer expectations and tighter assessment alignment.

Critical thinking displayed the greatest variability across pilots. The very large gains observed in UPATRAS contrast with moderate gains in IASIS and declines in EASD. From a theoretical standpoint, critical thinking is closely tied to pedagogical designs that emphasise problem identification, exploration, and reflective analysis. The findings suggest that when pilot implementations explicitly scaffold these processes, substantial improvements are possible, particularly when pre-pilot baselines are relatively low. Conversely, when pilot conditions prioritise task completion or delivery efficiency over epistemic exploration, gains in critical thinking may be constrained or even reversed. UPATRAS pilot serves as a concrete illustration of the design sensitivity in the evolution of critical thinking. In the context of UPATRAS, learning activities placed a strong emphasis on problem-driven tasks that required learners to analyse complex scenarios, justify their decisions, and iteratively refine their solutions based on feedback. Assessment criteria explicitly prioritised argumentation quality, evidence use, and reflective reasoning. This alignment between task design, assessment focus, and epistemic demands likely contributed to the substantial gains observed in critical thinking, particularly given the relatively low pre-pilot baseline.

Creativity outcomes were markedly context-dependent. While moderate to strong gains were observed in IASIS and UPATRAS, EASD experienced a decline from pre-pilot to pilot phases. Creativity is commonly understood as emerging from a balance between novelty, effectiveness, and implementability. Pilot phases that impose tighter schedules, predefined outputs, or stronger assessment constraints may inadvertently reduce opportunities for divergent thinking and exploratory problem-solving. The observed variability reinforces the view that creativity is particularly sensitive to instructional flexibility and the availability of open-ended learning tasks but also to the potential of being creative over time, showing not only a creative intention but creative perseverance (Leroy & Romero, 2022).

Collaboration emerged as the most fragile of the four competencies. Two pilots showed declines, while the third demonstrated only marginal stability. This pattern is consistent with theoretical and empirical work suggesting that collaboration depends heavily on sustained interaction, reciprocity, and carefully designed group structures. As learning initiatives scale from pre-pilot to pilot phases, reductions in synchronous interaction, changes in group composition, or increased individual accountability may weaken collaborative dynamics, even when collaboration remains an explicit learning objective.

## 5.2. Cross-Context Variability and Design Sensitivity

The divergent trajectories observed across pilots also highlight the importance of contextual factors. Differences in pedagogical emphasis, task design, and implementation strategies appear to play a significant role in shaping competency outcomes. The strong gains in critical thinking observed in UPATRAS, for example, suggest that targeted design choices can yield substantial benefits, while the declines observed in EASD indicate that competency development may be compromised when higher-order or collaborative elements are deprioritised. These variations should not be interpreted as reflecting differences in learner ability. Rather, they point to the central role of learning design, assessment practices, and activity structures in enabling or constraining the development of specific competencies. From this perspective, declines or stagnation in certain competencies are better understood as structural effects of implementation choices rather than as failures of learners or educators.

# 6. Conclusions

## 6.1. Limitations

Several limitations should be acknowledged. The quasi-experimental design does not allow for causal inference, and observed changes in competencies cannot be attributed solely to pilot interventions. Furthermore, collaboration outcomes may be under-represented in grade-based analyses, as assessment practices primarily emphasised individual performance. Variability in data availability across pilots, particularly the absence of communication data for UPATRAS, also limits full cross-context comparability.

## 6.2. Implications for Competency-Based Learning Design and Research

Our study findings underscore that improvements in transversal competencies are not automatic outcomes of scaling educational interventions. Instead, each competency responds to distinct pedagogical and assessment conditions. Competencies that are closely aligned with individual performance and assessment criteria, such as communication and aspects of critical thinking, appear more resilient during pilot scaling. In contrast, competencies that rely on social interaction or creative exploration are more vulnerable to structural constraints. These results support the argument that transversal competencies should not be treated as a homogeneous construct (Voogt & Roblin, 2012). Aggregating the 4Cs into a single composite indicator risks obscuring important differences in how competencies develop and how they are affected by design decisions. A competency-sensitive approach to both instructional design and evaluation is therefore essential, particularly in modular learning environments where learning activities and assessment practices may vary substantially across modules.

# Acknowledgements

This study is developed with the support of the Horizon augMENTOR project. Grant agreement ID: 101061509. DOI 10.3030/101061509.

**Conflicts of Interest**

The author declares no conflicts of interest regarding the publication of this paper.